\documentclass[prb,twocolumn,superscriptaddress,floatfix,citeautoscript]{revtex4}
\usepackage{amssymb,latexsym}
\usepackage{amsmath}    
\DeclareMathOperator{\sgn}{sgn}
\usepackage{graphicx}   
\usepackage{verbatim}   
\usepackage{color}      
\usepackage{subfigure}  
\usepackage{hyperref}   
\newcommand\figref{Fig.\ \ref} 
\usepackage{nicefrac}
\usepackage{multirow}	

\begin{document}
\title{Electron Scattering in 2D Semiconductors: Contrasting Dirac and Schr\"odinger Behavior}

\author{D. Meneses-Gustin}
\affiliation{Departamento de F\'{\i}sica, Universidade Federal de S\~{a}o Carlos, 13565-905 S\~{a}o Carlos, SP, Brazil}
\author{S. E. Ulloa}
\affiliation{Department of Physics and Astronomy and Nanoscale and Quantum Phenomena Institute, Ohio University, Athens, OH 45701-2979, USA}
\author{V. Lopez-Richard}
\affiliation{Departamento de F\'{\i}sica, Universidade Federal de S\~{a}o Carlos, 13565-905 S\~{a}o Carlos, SP, Brazil}

\begin{abstract}
Electronic transport through a material depends on the response to local perturbations induced by defects or impurities in the material. The scattering processes can be described in terms of phase shifts and corresponding cross sections. The multiorbital nature of the spinor states in transition metal dichalcogenides would naturally suggest the consideration of a massive Dirac equation to describe the problem, while the parabolic dispersion of its conduction and valence bands would invite a simpler Schr\"odinger equation description. Here, we contrast the scattering of massive Dirac particles and Schr\"odinger electrons, in order to assess different asymptotic regimes (low and high Fermi energy) for each one of the electronic models and describe their regime of validity or transition. At low energies, where the dispersion is approximately parabolic, the scattering processes are dominated by low angular momentum channels, which results in nearly isotropic scattering amplitudes. On the other hand, the differential cross section at high Fermi energies exhibits clear signatures of the linear band dispersion, as the partial phase shifts approach a non-zero value. We analyze the electronic dynamics by presenting differential cross sections for both attractive and repulsive scattering centers.
The dissimilar behavior between Dirac and Schr\"odinger carriers points to the limits and conditions over which different descriptions are required for the
reliable treatment of scattering processes in these materials.
\end{abstract}

\date{\today}
\maketitle

\section{Introduction}

Transition metal dichalcogenides (TMDs) are materials with atoms arranged in a layered X-M-X structure, where M is a transition metal, such as Mo and W, and X = S, Se, Te, a chalcogen. While atoms of the same trilayer (or ``monolayer'') are covalently bonded, the interactions between trilayers take place through weaker van der Waals forces \cite{Geim+}. These materials exhibit drastic changes in their electronic band structure upon easy exfoliation \cite{Cao2012a}. In MoS$_2$, for example, an electro-luminescent response appears in the visible range up to room temperature, \cite{Sercombe2013,Kolobov2016} as a consequence of the direct band gap of its monolayer form. The structural flexibility and semiconducting behavior of these two-dimensional TMDs makes them suitable for a host of novel optoelectronics applications \cite{Komsa2013, Wilson1969a, Castellanos-Gomez2010a}. Moreover, the study of MoS$_2$ and related TMDs poses a number of fundamental physical questions of interest which enrich our knowledge of quantum properties of condensed matter. \cite{QMat}

One interesting feature of semiconducting TMDs is that the low energy dynamics of their charge carriers is typically described by a minimal two-band model that can be seen as a massive Dirac Hamiltonian in two spatial dimensions.\cite{Xiao2012}  This formulation takes into account lattice symmetries, as well as  spin and valley degrees of freedom.  The Hamiltonian and corresponding eigenstates incorporate Berry phase structure which is especially interesting and important near the two direct-gap inequivalent valleys in reciprocal space. \cite{NiuRMP}  As a consequence, the optical response of the material encodes spin-valley coupling features of the state, which have given rise to long-lived polarization memory at low temperatures,\cite{Xu} as well as interesting spin-polarized optically-produced electronic currents.\cite{Xiao2012}  The Dirac character of the carriers in these materials would also give rise to different Landau level semiclassical phases, suitable for observation in Shubnikov-de Hass magnetotransport oscillations. \cite{Goerbig2013}

The two-dimensional nature of monolayer TMD systems makes them susceptible to effects introduced by the substrates as well as intentional or ambient contamination, which can introduce charged defects/impurities or local changes in the doping level \cite{Buscema2014,Feng2012a} that would be expected to modulate the transport response.  An interesting example of this behavior was reported recently, \cite{Margapoti2014} as the deposition of photoresponsive azobenzene molecules was shown to produce changes in the optical and transport properties. As these and similar perturbations can be seen as local disturbances to the corresponding Hamiltonian of the system, it is important to study how they affect the electronic dynamics in these materials.

Here, we introduce local perturbations on the 2D electronic dynamics and discuss the resulting scattering problem for carriers described by a massive Dirac formalism.  We describe the electronic transport problem through an analysis of the differential scattering cross section and compare these results with those expected for a simple parabolic-mass dispersion, as would be typically used in semiconducting systems. The comparison of the two approaches provides signatures of the peculiar Dirac spinor structure which can be quantified through differences in anisotropic scattering and its energy dependence.
The two-dimensional scattering problem is characterized in terms of the scattering phase shifts and differential cross section and their dependence on externally controllable parameters. The conditions for the validity of a parabolic or full Dirac dispersion relation are analyzed in detail, especially by contrasting asymptotic limits of scattering amplitudes, and the different features of their scattering angle dependence.  The details of the different functional behavior would have direct impact on the resulting transport and elastic times, and correspondingly different transport experiments.

\section{TMD Hamiltonian and Scattering amplitude}
The Hamiltonian describing the dynamics near the K-valleys in the Brillouin zone of the TMD monolayer can be written in an effective massive Dirac form
\cite{Xiao2012}
\begin{align}\label{eq:Hamiltonian}
H_D = at (\tau q_x \sigma_x +q_y \sigma_y) + \frac{\Delta}{2}\sigma_z - \tau \lambda \frac{\sigma_z-1}{2}s_z ,
\end{align}
for small momenta $\mathbf{q} = (q_x, q_y)$ about the K$_\tau$ valleys, $\tau =\pm1$ (known as the K and K' valleys, respectively),
where $a$ is the lattice constant, $t$ the effective hopping integral,  $\sigma_i$ the Pauli matrices that act on the degree of freedom corresponding to the two main $d$-orbitals of transition metal atoms in the hexagonal lattice (pseudospin), $\Delta$ plays the role of the rest mass, and $2\lambda$ is the spin splitting at the valence band due to spin-orbit interaction, and $s_z$ is the Pauli matrix for spin.  We note that additional quadratic terms may appear in the effective Hamiltonian \cite{Guido,Carlos}, resulting in qualitatively similar scattering behavior in what follows.

Eigenstates of this Hamiltonian can be described as plane wave spinor states with momentum K$_\tau + \mathbf{q}$.  However, the addition of a perturbing potential $V(\mathbf{r})$ gives rise to more complex solutions and corresponding density distribution profiles.  We will describe the scattering amplitude $f(\theta)$ arising from such a potential, under the assumption of circular symmetry and smooth radial direction dependence, so that intervalley scattering is negligible.  One should comment that lattice parity and time reversal symmetries impose restrictions on the scattering amplitude, similar to what has been described for graphene systems, where a similar (but massless) Hamiltonian describes the carrier dynamics. \cite{Asmar2015}

\begin{figure}[ht]
\centering
 \includegraphics[scale=0.8]{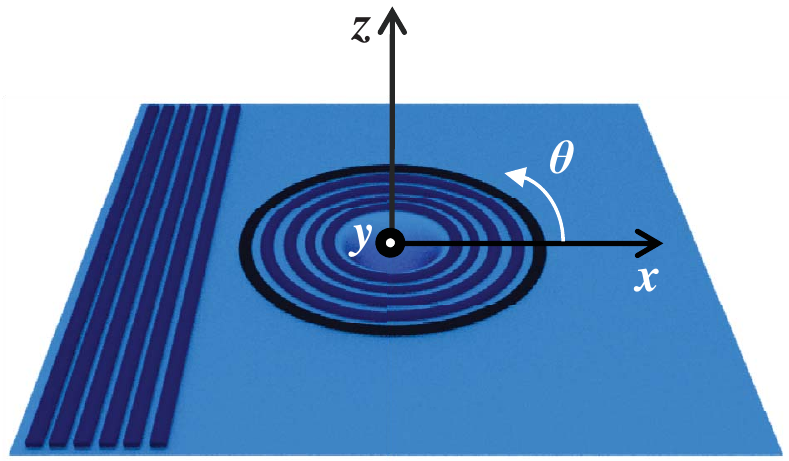}
 \caption{The scattering by a central potential of incident planes waves from the left is represented with outgoing circular waves produced by scattering. The angle of detection $\theta$ is also indicated with $x$-$y$ defining the plane of dynamics.}\label{fig:well_barrier}
\end{figure}

We consider a scalar scattering center placed at the origin of coordinates, as represented in \figref{fig:well_barrier}, and characterized by a constant $V(\mathbf{r})=V$  within a radius $L$, while vanishing elsewhere.
As we assume vanishingly weak intervalley scattering that is independent of spin, the otherwise $8\times8$ spinor problem decomposes into four $2\times2$ block diagonal problems.  In the following we focus on the K ($\tau=1$) valley, and suppress the spin index in favor of the implicit sign of the spin-orbit coupling $\lambda$.

We look for solutions $\psi$ satisfying simultaneously
\begin{align}
 \left[H_D + V(\mathbf{r})\right]\psi =& E\psi, \label{eq:simultaneoulyDiraca}\\
  \psi (r \to \infty) \longrightarrow &
\psi_{q}  e^{i \mathbf{q} \cdot \mathbf{r}} + f_q( \theta)
\psi_{q} \frac{e^{i q r}}{\sqrt{r}},\label{eq:simultaneoulyDiracb}
 \end{align}
where $\mathbf{q}=q( \cos \theta_q, \sin \theta_q)$ is the wave vector of the incident plane wave spinor at energy $E$, with
$\psi_{q} = \left(1, \frac{- at q(E)}{\nicefrac{\lambda -\Delta}{2}  - E} e^{i \theta_q}\right)^T$, where
\begin{equation} \label{eq:q}
q (E) = \frac{1}{at} \left(E^2 - \lambda E - \frac{\Delta}{2} \left(\frac{\Delta}{2} -\lambda\right)\right)^{1/2} \, ,
\end{equation}
and for real $q$ one requires $E\geq E_c \equiv \frac{\Delta}{2}$ or $E \leq E_v \equiv \lambda - \frac{\Delta}{2}$.
The first term in Eq. \eqref{eq:simultaneoulyDiracb},  is the incident spinor, while the second term represents a circular wave leaving the potential region,
with a scattering amplitude $f_q(\theta)$, where $\theta$ is the angular coordinate measured with respect to the incident direction, see \figref{fig:well_barrier}.
The differential cross section is then defined by
\begin{align}\label{eq:f2}
\frac{d\sigma}{d\theta} = |f_q(\theta)|^2.
\end{align}

We exploit the circular symmetry of the problem and decompose the scattering amplitude in angular momentum channels with index $m$,
\begin{align}\label{eq:f2Dirac}
 f_q(\theta) = \sum_{m=-\infty}^\infty \frac{1}{2}(e^{2i \delta_m} - 1)\sqrt{\frac{2}{i \pi q}}e^{i m \theta} \, ,
\end{align}
written in terms of the corresponding phase shifts $\delta_m$, generated by the interaction with the potential, and modifying the incident wave.

\begin{figure}[ht]
\centering
 \includegraphics[scale=0.175]{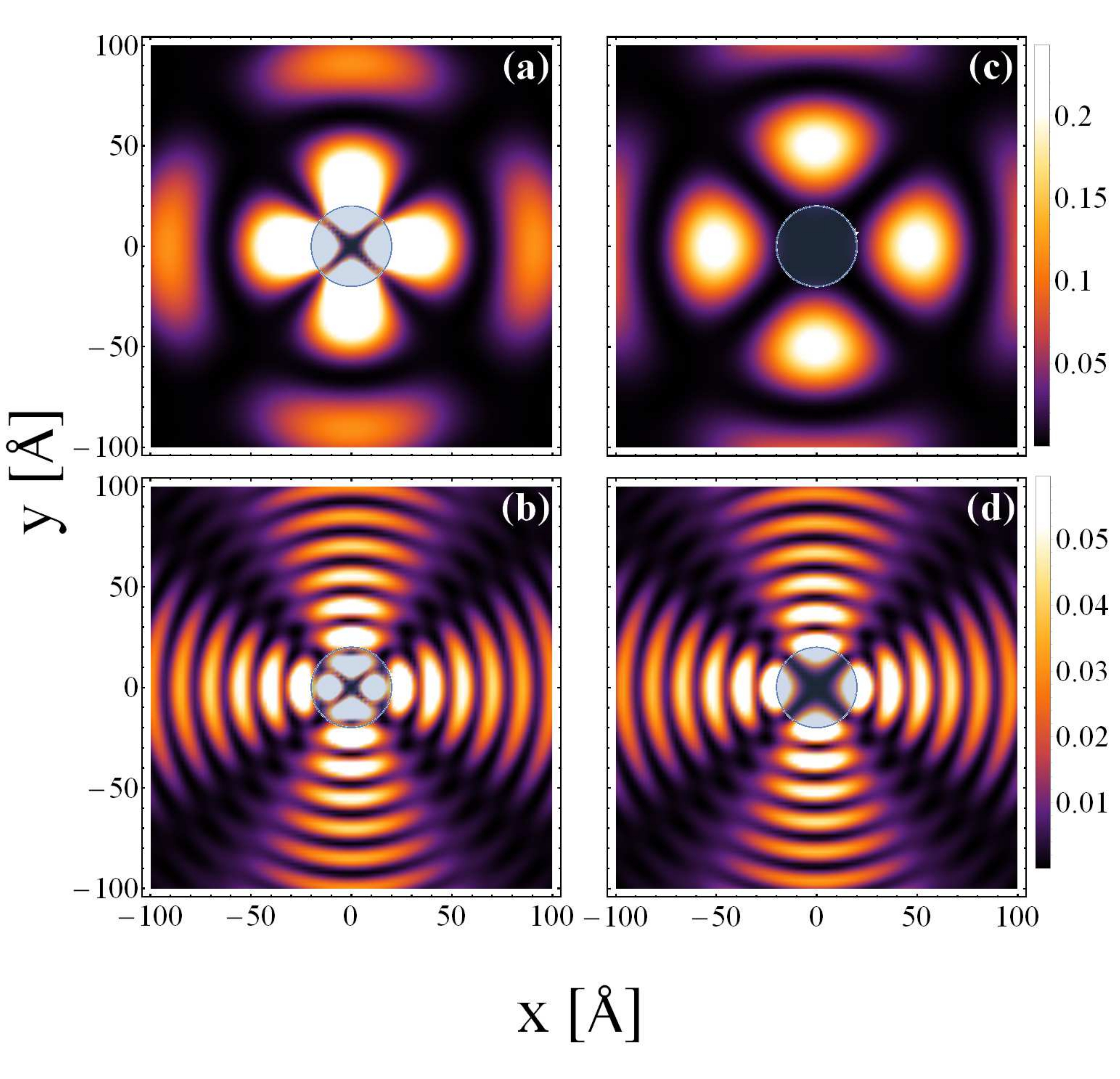}
 \caption{Scattered wave profile, $|{\rm Re} \, \psi_{\rm scatt}|^2$, for angular momentum $m=2$ for massive Dirac carriers. The patterns correspond to a particle scattered by a potential well problem with (a) low, $0.03$eV, and (b) high, $0.3$eV, kinetic energy respectively. (c) and (d) show similar probability density for a barrier potential, also for the same two energy values.} \label{fig:psi2}
\end{figure}

The phase shifts can be obtained from the continuity of the spinors at the edge of the potential, $\psi_m^i\big|_L = \psi_m^o\big| _L $, where $i, o$ label the region inside and outside of the potential, respectively. This condition yields
\begin{align}
 e^{2i \delta_m} &= \frac{H^{(2)}_{m+1}(qL)J_m(\tilde{q}L)- DH^{(2)}_m(qL)J_{m+1}(\tilde{q}L)}{DH^{(1)}_{m}(qL)J_{m+1}(\tilde{q}L) -H^{(1)}_{m+1}(qL)J_m(\tilde{q}L)} \label{eq:phaseDirac},
\end{align}
with
\begin{equation}
D=\frac{\nicefrac{-\Delta}{2}+ \lambda - E}{\nicefrac{-\Delta}{2}+ \lambda - (E-V)}\frac{\tilde{q}}{q} \, ,
\end{equation}
and where $\tilde{q}(E) \equiv q(E-V)$ is the wave vector inside the potential region.   The scattered probability density can then be calculated; an example for $m=2$ is shown in \figref{fig:psi2}. From this plot of the scattered wave is clear than a higher transmission probability is inside the potential well (left column), as one would expect, and with slightly broader maxima than in the repulsive barrier (right column).

In order to assess the charge transport response to perturbations in TMD systems, we obtain the scattering amplitude and phase shifts for different parameters of the potential (depth and width), and examine their energy dependence. As we will see below, at low energies the problem can be described in terms of an effective mass in the Schr\"odinger description, as one would anticipate.  However, the Dirac description has strong deviations from the effective mass approach even at moderate energies.  This point is interesting since the Schr\"odinger equation remains the natural theoretical path to describe quantum effects in many semiconductor systems.

\section{Schr\"odinger limit at low energies}

For low wave vectors, one reasonable approximation for the electronic structure of TMDs would be a parabolic dispersion \cite{Goerbig2013}. Such limit can be  made evident from Eq.\  \eqref{eq:q}, assuming $\Delta > \lambda$ and small $q$, so that
\begin{align}
E(q \rightarrow 0)  &\thickapprox
\begin{cases}
\frac{\Delta}{2} + \frac{(at)^2 }{\Delta - \lambda}q^2, \\
\lambda -\frac{\Delta}{2}  - \frac{(at)^2 }{\Delta - \lambda}q^2,
\end{cases}
\end{align}
where clearly $m^* = \pm {\hbar^2(\Delta - \lambda)}/{2a^2t^2}$ is the effective mass for electrons (+) or holes ($-$). [As mentioned above, other quadratic terms may also contribute to the TMD effective masses,\cite{Guido, Carlos} although they do not affect our discussion and main conclusions below.]

In the Schr\"odinger parabolic dispersion, one can of course determine the scattering amplitude in terms of the corresponding phase shifts, as obtained from the appropriate continuity conditions, so that \cite{QMbook}
\begin{align}\label{eq:phaseSch}
e^{2i \delta_m^S} = \frac{H^{(2)'}_{m}(k L)- B H^{(2)}_m(k L)}{B H^{(1)}_{m}(k L)-H^{(1)'}_m(k L)},
\end{align}
where $B = \frac{\tilde{k} J_m'(\tilde{k} L)}{k  J_m(\tilde{k} L)}$, and use $k$ as the wave number in the Schr\"odinger description, $k(E)=\sqrt{2m^* E/\hbar^2}$. As before, we define $\tilde{k}\equiv k(E-V)$, and have added a superscript to $\delta_m^S$ as a reminder this is the phase shift for the parabolic dispersion.  A comparison of Eqs.\ \eqref{eq:phaseSch} and \eqref{eq:phaseDirac} makes it clear that the phase shifts contain intrinsically different information on the 2D dynamics of these different spinors.  One would anticipate this difference to be reflected in the differential cross sections, as we see below.

\section{Energy dependence of Phase shifts}

Based on the expressions above for the different phase shifts, we proceed to compare asymptotic behaviors of the scattering problem in the Dirac or Schr\"odinger formulations.  It is useful to rewrite the phase shift expression \eqref{eq:phaseDirac} as
\begingroup
\allowdisplaybreaks
\begin{align}\label{eq:tanDirac}
 \tan \delta_m &= \frac{J_{m+1}(qL)J_m(\tilde{q}L) - DJ_{m+1}(\tilde{q}L)J_m(qL)}{Y_{m+1}(qL)J_m(\tilde{q}L) - DJ_{m+1}(\tilde{q}L)Y_m(qL)} \, ;
\end{align}
\endgroup
notice that each phase shift is undetermined up to a multiple of $\pi$.

We analyze first the short wavelength (high energy) limit, $q \to \infty$, such that $D \approx 1 $.
Note that in the high energy limit, $q \approx \tilde{q}$, however since the dispersion relation is hyperbolic, the wave vector proportional to energy, which results in a non-vanishing Dirac phase shift $\delta_m$, \cite{Novikov,Asmar2014}
\begin{align}\label{eq:DiracInfty}
 \lim_{q \to \infty} \delta_m &= -\frac{LV}{at}.
\end{align}
This result reflects an interesting counterintuitive insight: incoming electrons \emph{see} the potential even when moving at high energy. This means that in a TMD material with electronic structure
described by the 2D Dirac Hamiltonian, the carrier mobility may be strongly affected by perturbations created on its surface, even at high energies.

We now analyze the long wavenumber (low energy) phase shift, which requires careful consideration of the sign of the potential. Notice that one should consider both the well and barrier situations se\-pa\-ra\-te\-ly. For $V < E$, the $J_m(\tilde{q}r)$ and $Y_m(\tilde{q}r)$ functions are convenient solutions to the Bessel equation, whereas for a quantum barrier at low energy, $V>E$, the wave vector $q$, and consequently $D$, become complex. The solutions are the modified Bessel functions of the first kind, $I_s$ (finite at the origin), and the second kind, $K_s$ (singular at $x = 0$). These considerations result in
\begin{align}\label{eq:Dirac0}
\lim_{q \to 0} \delta_m =
\begin{cases}
 -\frac{\pi m}{m!^2}\left(\frac{qL}{2}\right)^{2m} \sgn(V), & \text{for } m \neq 0,\\
\frac{\pi}{2} \frac{1}{\ln \frac{qL}{2} + \gamma},  & \text{for } m = 0.
\end{cases}
\end{align}
This limit shows that the main contributing channel in the scattering process at low energy is $m=0$, with higher angular momenta contributing only weakly.

For parabolic band carriers, Eq.\ \eqref{eq:phaseSch} can also be cast as
\begingroup
\allowdisplaybreaks
\begin{align}
 \tan \delta_m^S &= \frac{J'_m(kL) - BJ_m(kL)}{Y'_m(kL)  - BY_m(kL)}.
\label{eq:taS}
\end{align}
\endgroup
For high scattering energy this yields
\begin{align}\label{eq:SCHInfty}
 \lim_{k \to \infty} \delta_m^S & = - \frac{2m^*}{\hbar^2} \frac{VL}{2k},
\end{align}
which is substantially different from the corresponding expression \eqref{eq:DiracInfty} for Dirac scattering. Most importantly, as the wave vector grows the carriers are scattered less by the perturbation in the Schr\"odinger formulation, with phase shifts vanishing asymptotically.

In the low energy regime we have $\tilde{k} =\sqrt{2m^* (E-V)/\hbar ^2} $, which becomes imaginary for energies below the barrier.  The
corresponding small wave number argument for $J_m$ and $Y_m$ results in
\begin{align}\label{eq:SCH0}
\lim_{k \to 0} \delta_m^S =
\begin{cases}
 -\frac{\pi m}{m!^2}\left(\frac{kL}{2}\right)^{2m} \sgn(V), & \text{for } m \neq 0, \\
 \frac{\pi}{2}\frac{1}{\ln\frac{kL}{2}+ \gamma } & \text{for } m = 0,
\end{cases}
\end{align}
which is identical to the corresponding expression for Dirac carriers in \eqref{eq:Dirac0}. The scattering phase shifts for electrons moving at low energy vanish, as the wavelength is much smaller than the size of the perturbation $kL \ll 1$.

\begin{figure}[ht]
\centering
 \includegraphics[scale=0.4]{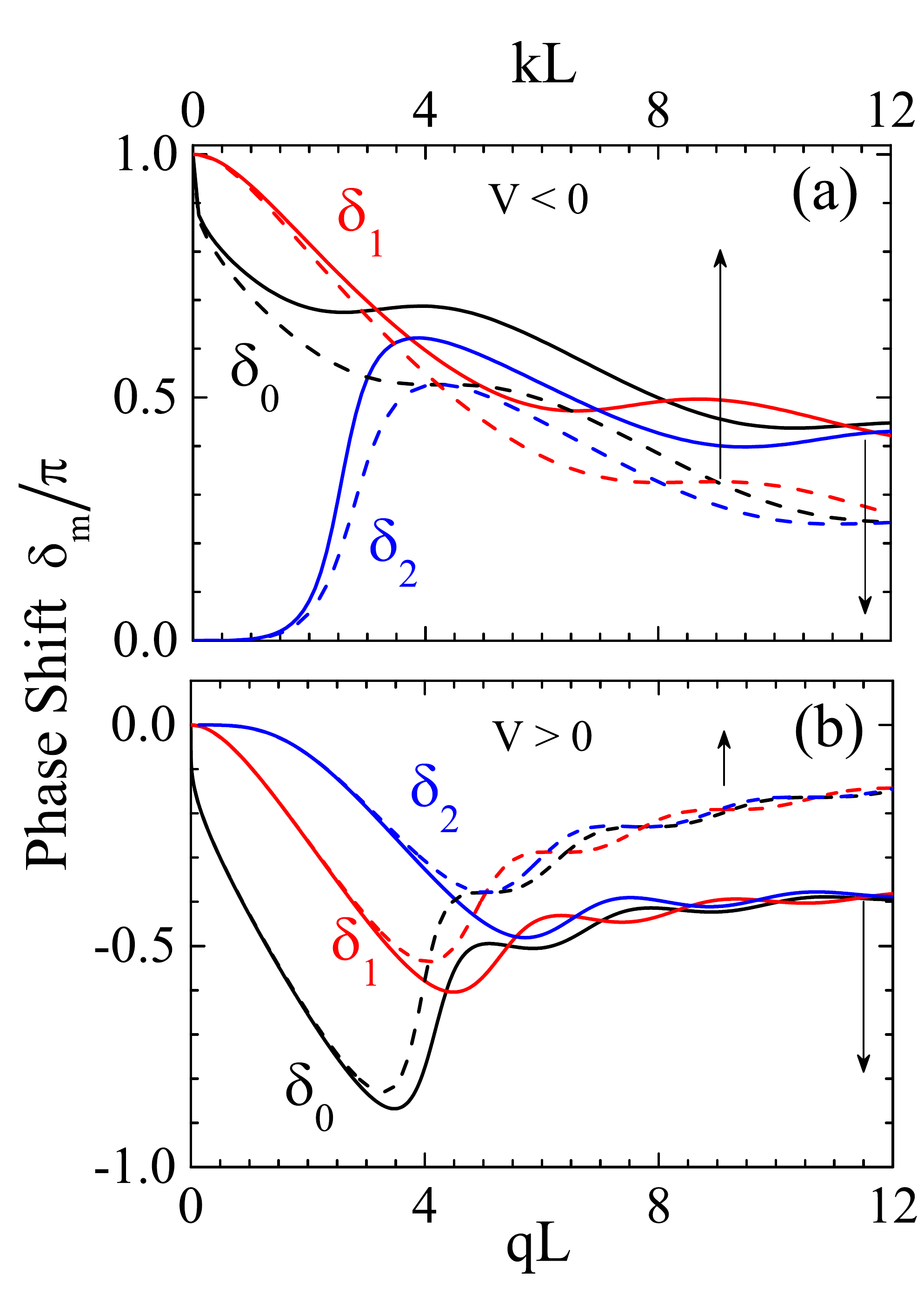}
 \caption{Energy dependence of scattering phase shifts for Dirac (continuous lines) and Schr\"odinger dispersion (dashed lines). The potential energy is (a) attractive, $V=-0.2$eV, and (b) repulsive, $V=0.2$eV, respectively. The width $L=20$\AA\ is kept fixed.  Noticeable differences appear at lower energies for attractive potentials; asymptotic values vanish for Schr\"odinger particles, while they have a finite value for Dirac, as in Eq.\ \eqref{eq:DiracInfty}.} \label{fig:ASYM}
\end{figure}

The differences and similarities between Dirac and Schr\"odinger formulations found above become perhaps more noticeable in graphics. In \figref{fig:ASYM} we plot both phase shifts $\delta_m$ and $\delta_m^S$ as function of the wave vector for a fixed target radius $L=20$\AA.
For concreteness we consider a MoS$_2$ monolayer, with parameters found in Ref.\ [\onlinecite{Xiao2012}]: $a=3.193$\AA, $t=1.10$eV, $\Delta=1.66$eV and $2\lambda=0.15$eV.
As expressed by \eqref{eq:Dirac0} and \eqref{eq:SCH0}, both formulations match at short wave vectors, with Dirac (Schr\"odinger) phase shifts shown with continuous (dashed) lines.
Phase shifts are seen to differ for $qL \gtrsim 3$, and even earlier; this is especially true for the case of a potential well, as seen in \figref{fig:ASYM}(a).  Notice also that the low-energy
limit of $\delta_m$ is not necessarily zero, as we see for $d_j$, for $j=0,1$ in the attractive potential case.  From this behavior,
it is possible to extract information about the number of bound states with angular momentum $m$, $n_m$.
For a short-range attractive potential, Levinson's theorem \cite{PhysRevA.57.3478} shows that a finite number of bound states appear if the potential is strong enough,
as reflected in the corresponding phase shift taking the value $n_m \pi$ at low energy, whereas for a repulsive potential one finds $n_m=0$ always.

For high energies, the Dirac-like dynamics results in finite phase shifts, clearly seen in both panels of \figref{fig:ASYM}.  This is in contrast to the case of Schr\"odinger-like carriers with high energy that exhibit vanishingly small phase shifts. These differences would, of course, be evident in the resulting cross sections, as we study next.

\section{Comparison of Differential cross sections}

\begin{figure}[h]
\centering
 \includegraphics[scale=0.3]{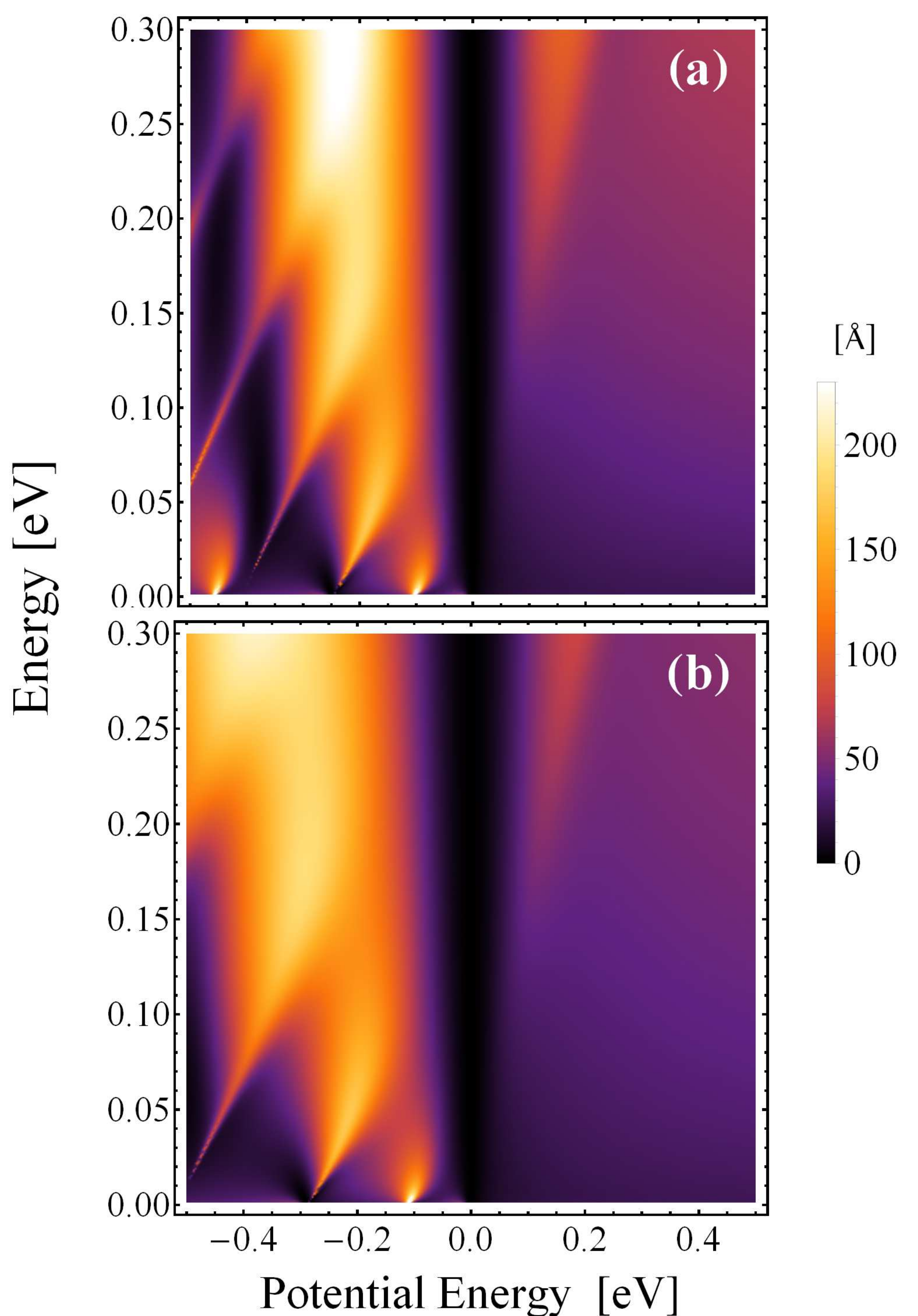}
 \caption{Forward differential scattering cross section, $| f_q(\theta = 0) |^2$ as function of potential $V$ and incident kinetic energy. (a) For massive Dirac Hamiltonian, and (b) for parabolic  dispersion relation. $L = 20$\AA.} \label{fig:f2vsVvsE}
\end{figure}

 Different features for electronic transport appear by using repulsive ($V>0$) or attractive ($V<0$) potentials as scattering centers, as could be produced in TMD systems. This is evident in \figref{fig:f2vsVvsE}(a), where a map of the differential cross section for forward scattering ($\theta = 0$) is shown as potential and scattering energy are varied. Although the qualitative scattering features of Dirac-massive carriers are reproduced by the Schr\"odinger description for an appropriate effective mass, displayed in \figref{fig:f2vsVvsE}(b), clear quantitative differences emerge.

In both cases we notice the presence of resonant enhancement of the forward scattering for attractive potentials, with overall higher amplitudes in the Dirac case.  It is also clear that the
resonance separation is smaller for Dirac than for Schr\"odinger particles, as one could naively expect from wavelength quantization arguments in both descriptions.
Notice also that for
repulsive potentials the energy dependence is rather monotonic, with a slowly increasing amplitude for higher energies. As expected, both panels in \figref{fig:f2vsVvsE} show similar features at low energy, and differ more strongly as the energy increases.

\begin{figure}[h]
\centering
 \includegraphics[scale=0.18]{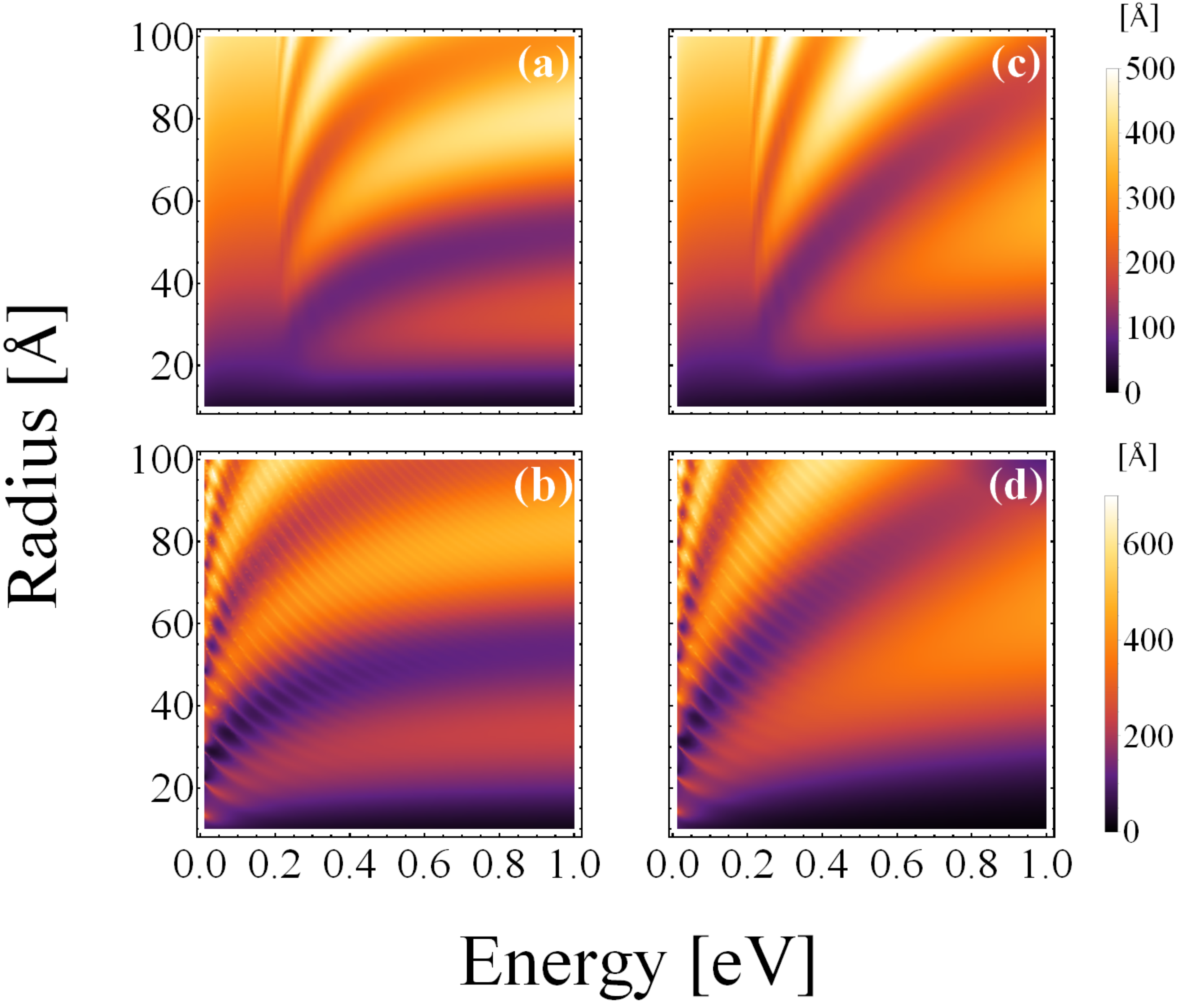}
 \caption{Total cross section, calculated as in Eq. (\ref{eq:totsigma}), as a function of the incident energy for (a) $V=0.2$eV and (b) $V=-0.2$eV, respectively. (c) and (d) show the same as (a) and (b) but using the parabolic mass approximation.}\label{fig:fvsR}
\end{figure}

Let us now examine the total scattering cross section, obtained after angular integration of the differential cross section,
\begin{equation}\label{eq:totsigma}
\sigma_{\rm tot} = \int d \theta \, \frac{d\sigma}{d\theta} \, ,
\end{equation}
as the radius of the perturbation potential changes.
The Dirac-massive results are shown in \figref{fig:fvsR}(a) for $V>0$ and in \ref{fig:fvsR}(b) for $V<0$.  The results for Schr\"odinger carriers are shown in \figref{fig:fvsR}(c) and (d).
Comparing different panels, it is clear that the scattering probability for a Dirac particle has a weaker energy dependence than in the Schr\"odinger case. In the latter, the scattering cross section depends strongly on energy, even for high energy values and for both signs of $V$. There is a clear boundary for repulsive potentials in both models, as the scattering amplitude varies smoothly for $E<V$, before resonances appear for $E>V$. The oscillatory resonance patterns reflect the interference of transmitted and reflected waves from the geometric boundaries of the scattering center.

\begin{figure}[h]
\centering
 \includegraphics[scale=0.4]{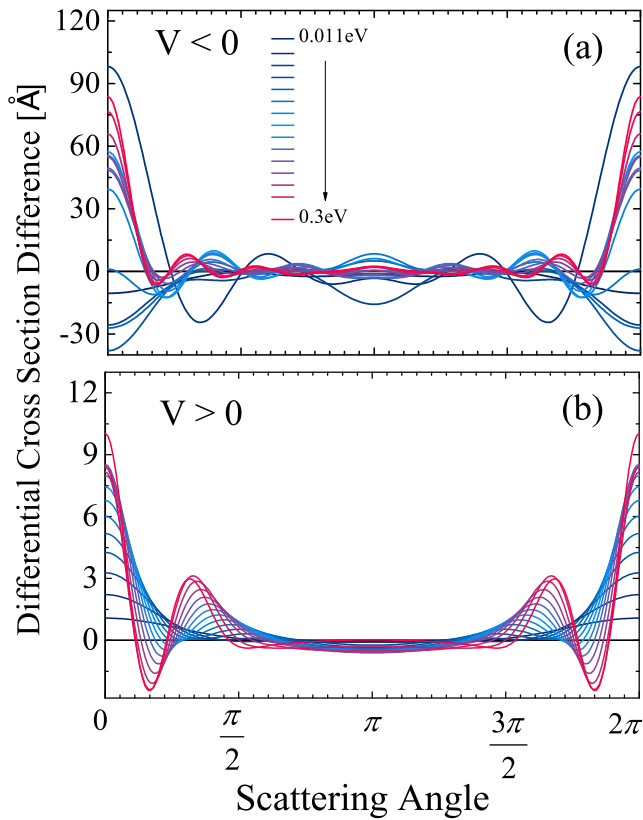}
 \caption{The contrast in differential cross sections, $\Delta \sigma =\frac{d\sigma^{D}}{d\theta}-\frac{d\sigma^{S}}{d\theta}$, is plotted as function of the scattering angle $\theta$ and for different kinetic energy and potential values. (a) Results for attractive potential ($V<0$);  and (b) for repulsive barrier ($V>0$).  Radius is kept at $L=20$ \AA.}\label{fig:DIFF}
\end{figure}

In order to quantitatively assess and exhibit the difference of scattering amplitudes in the Dirac and the Schr\"odinger descriptions, we plot in \figref{fig:DIFF} the difference in differential cross sections, $\Delta \sigma = \frac{d\sigma^{D}}{d\theta}-\frac{d\sigma^{S}}{d\theta}$, as function of the scattering angle for different energy of incident carriers.  The negative values in these plots indicate a greater probability for scattering in the case of Schr\"odinger carriers, as we see in \figref{fig:DIFF} for different signs of $V$. The main difference appears in the forward scattering direction for both positive and negative potentials. Notice the contrast grows monotonically with energy in the repulsive case ($V>0$, panel (b)).  However, the contrast for the attractive case has a non-monotonic energy dependence. This behavior also extends over the entire scattering angle range and exhibits strong angular dependence.   The variations seen in \figref{fig:DIFF} as energy grows have clear enhanced anisotropy, which would affect observables in electronic transport experiments.  Notice moreover that both descriptions result in very similar non-zero backscattering amplitudes, a result of the finite mass in the Dirac dispersion which negates the appearance of Klein tunneling. \cite{Asmar2014}

\section{Conclusions}
In summary, we have used scattering theory in two dimensions to find the differential cross section generated by both repulsive and attractive potentials in an assumed circular symmetry.
We have also explored the modulation of scattering for different potential features, as well as different electron energies.  A comparison between Dirac and parabolic-band Schr\"odinger carriers in terms of scattering phase shifts demonstrates that both descriptions are comparable at low incident energy, as expected.  However, for high energies we find that carriers ruled by the Dirac equation have higher probability to be found in the forward direction, which could be seen as a remnant of the Klein paradox found in the Dirac massless case \cite{Klein}.  \\

\acknowledgments The authors are grateful for the financial support from the Brazilian Agencies FAPESP (grants $2013/24253-5$, $2016/02065-0$, and $2014/02112-3$), CNPq (grant $306414/2015-5$), and NSF-DMR grant 1508325 (Ohio).


\begin{thebibliography}{0}
\expandafter\ifx\csname natexlab\endcsname\relax\def\natexlab#1{#1}\fi
\expandafter\ifx\csname bibnamefont\endcsname\relax
  \def\bibnamefont#1{#1}\fi
\expandafter\ifx\csname bibfnamefont\endcsname\relax
  \def\bibfnamefont#1{#1}\fi
\expandafter\ifx\csname citenamefont\endcsname\relax
  \def\citenamefont#1{#1}\fi
\expandafter\ifx\csname url\endcsname\relax
  \def\url#1{\texttt{#1}}\fi
\expandafter\ifx\csname urlprefix\endcsname\relax\def\urlprefix{URL }\fi
\providecommand{\bibinfo}[2]{#2}
\providecommand{\eprint}[2][]{\url{#2}}

\end{thebibliography}


\begin{thebibliography}{99}

\bibitem{Geim+} A. K. Geim and I. V. Grigorieva, Nature \textbf{499}, 419 (2013).

\bibitem{Cao2012a} T. Cao, G. Wang, W. Han, H. Ye, C. Zhu, J. Shi, Q. Niu, P. Tan, E. Wang, B. Liu, and J. Feng, Nat. Commun. \textbf{3}, 887 (2012).

\bibitem{Sercombe2013} D. Sercombe, S. Schwarz, O. D. Pozo-Zamudio, F. Liu, B. J. Robinson, E. A. Chekhovich, I. I. Tartakovskii, O. Kolosov, and A. I. Tartakovskii, Nat. Commun. \textbf{3}, 3489 (2013).

\bibitem{Kolobov2016} A. V. Kolobov and J. Tominaga, vol. 239 Springer Series in Materials Science (Springer International Publishing, Cham, 2016).

\bibitem{Komsa2013} H. P. Komsa and A. V. Krasheninnikov, Phys. Rev. B. \textbf{88}, 1 (2013).

\bibitem{Wilson1969a} J. A. Wilson and A. D. Yoffe, Adv. Phys. \textbf{18}, 193 (1969).

\bibitem{Castellanos-Gomez2010a} A. Castellanos-Gomez, N. Agrat and G. Rubio-Bollinger, Appl. Phy. Lett. \textbf{96}, 213116 (2010).

\bibitem{QMat} B. Keimer and J. E. Moore, Nat. Phys. \textbf{13}, 1045 (2017).

\bibitem{Xiao2012} D. Xiao, G-B Liu, W. Feng, X. Xu, and W. Yao, Phys. Rev. Lett \textbf{108}, 196802 (2012).

\bibitem{NiuRMP} D. Xiao, M-C Chang and Q. Niu, Rev. Mod. Phys. \textbf{82}, 1959 (2010).

\bibitem{Xu} X. Xu, W. Yao, D. Xiao and T. F. Heinz, Nat. Phys. \textbf{10}, 343 (2014).

\bibitem{Goerbig2013} M. O. Goerbig, G. Montambaux and F. Pi\'echon, EPL. \textbf{105}, 57005 (2014).

\bibitem{Buscema2014} M. Buscema, G. A. Steele, H. S. J. van der Zant and A. Castellanos-Gomez, Nano Res. \textbf{7}, 1 (2014).

\bibitem{Feng2012a} J. Feng, X. Qian, C-W. Huang and J. Li, Nat. Photonics. \textbf{6}, 866 (2012).

\bibitem{Margapoti2014} E. Margapoti, P. Strobel, M. M. Asmar, M. Seifert, J. Li, M. Sachsenhauser, {\"{O}}. Ceylan, C.-A. Palma, J. V. Barth, J. A. Garrido, A. Cattani-Scholz, S. E. Ulloa, and J. J. Finley, Nano Lett. \textbf{14}, 6823 (2014).

\bibitem{Guido} A. Korm\'anyos, V. Z\'olyomi, N. D. Drummond, P. Rakyta, G. Burkard, and V. I. Fal'ko, Phys. Rev. B \textbf{88}, 045416 (2013).

\bibitem{Carlos} C. Segarra, J. Planelles and E. E. Ulloa, Phys. Rev. B \textbf{93}, 085312 (2016).

\bibitem{Asmar2015} M. M. Asmar and S. E. Ulloa, Phys. Rev. B \textbf{91}, 165407 (2015).

\bibitem{QMbook} Schwabl, F., \emph{Quantum Mechanics} (Springer, Berlin, 1993)

\bibitem{Asmar2014} M. M. Asmar and S. E. Ulloa, Phys. Rev. Lett. \textbf{112}, 136602 (2014).

\bibitem{Novikov} D. S. Novikov, Phys. Rev. B \textbf{76}, 245435 (2007).

\bibitem{PhysRevA.57.3478} Q-G Lin, Phys. Rev. A. \textbf{57}, 3478 (1998).

\bibitem{Klein} M. I. Katsnelson, K. S. Novoselov and A. K. Geim, Nat. Phys. \textbf{12}, 620 (2006).

\end{thebibliography}

\end{document}